# High-T$_C$ superconductivity induced by magnetic interactions

College of Applied Sciences, Beijing University of Technology, Beijing, China, 100124*

Jiang Jinhuan

**Abstract:** In this paper, a microscopic theory of magnetic-interaction-induced pairing in superconductivity of high temperature superconductors (HTSC) was developed on the basis of four idealized assumptions: (1) only a small number of electrons(or holes) are involved in superconductivity, and its density is $n\delta^2$; (2) magnetic interactions between electron spins lead to superconductivity; (3) there are different electronic states, i.e., the on-site doubly-occupied electrons forming anti-ferromagnetic insulator states, the off-site doubly-occupied electrons forming superconducting states, the singly-occupied (spin up or down) electrons forming normal states and the empty states; (4) the average kinetic energy of electrons (or holes) complies with the equipartition theorem of energy. Based on these assumptions, an approximate effective Hamiltonian was suggested. A parabolic relation between $T_C$ and the doping concentration $\delta$ was found and thus the phase diagram for HTSC has been explained. It was also found that, $T_C$ is related to the anti-ferromagnetic interaction energy $J$ (or critical magnetic field $B_C$) and the degrees of freedom of electrons $i$. The $T_C$ values are thus calculated from this theory to be 92.8K for YBa$_2$Cu$_3$O$_{6.15}$, 40.3K for La$_2$CuO$_4$, and 58K for SmOF$_e$As, which are in good agreement with the experimental results of 92K, 40K, and 54K, respectively. It was estimated that, $T_C$ in the slab HTSC is higher than that in the bulk, and $T_C$ for SmOF$_e$As can be up to 116K.

## 1. Four idealized assumptions

In 1911, Onnes found near zero resistance in mercury at 4.2K [1]. The temperature, where the superconductor loses its resistance, is called the superconducting critical temperature $T_C$. In 1914, Onnes found the superconducting state could be destroyed by an external magnetic field [2]. When the external magnetic field is greater than the critical magnetic field $B_C$, superconductivity disappears. In 1957, Bardeen, Cooper and Schrieffer developed a microscopic theory for superconductivity which was called BCS theory [3]. According to the BCS theory, the electrons near the Fermi surface can pair up (called Cooper pairs), forming bound states via electron-phonon interactions, and thus lowering the energy of the system. The Cooper pairs can undergo Bose-Einstein condensation and make the material superconducting. So far, Cooper pairs are found to exist in all kinds of superconductors, and the breakdown of Cooper pairs is the main reason that destroys superconductivity [4]. The energy needed to break up a Cooper pair, i.e., the bonding energy, is the double of the BCS superconducting energy gap $\Delta$. This energy gap is an important energy scale in superconductors that can determine $T_C$. The critical temperature $T_C$ induced by electron-phonon interactions is not so high, usually no higher than 40K estimated by the BCS theory. Therefore, this type of superconductors is called the low temperature superconductors. In the following, we introduce the experimental and theoretical progress on the research of high temperature superconductors (HTSC).

In 1986, Bednorz and Müller found $T_C$ around 30K in the Ba-La-Cu-O system [5]. Soon after, higher than 40K $T_C$ was demonstrated to exist in copper-oxide materials by several independent groups in China, Japan and USA, and this rapidly increased to 163K with further research in this area. This type of superconductors is called copper-oxide high temperature superconductors. $T_C$ in

---

Fe-based superconductors was reported to be higher than 50K [6]. Copper oxides without doping are anti-ferromagnetic insulators and typical strongly correlated electron systems. A common phase diagram is that: HTSC materials with the low doping concentration $\delta$ are anti-ferromagnetic insulators, and superconductivity appears only when $\delta$ reaches a certain value; $T_C$ will increase with increasing $\delta$ before it reaches a maximum value $T_{C,max}$, and then it begins to decrease with further increasing $\delta$. The optimal $\delta$ corresponds to the maximum $T_{C,max}$, and the regimes before and after it are called the under-doped and over-doped regimes, respectively. The superconducting state in the hole-type anti-ferromagnetic insulators appears at $\delta$ around 5%, and the optimal $\delta$ is around 16%. An empirical formula for HTSC phase diagram is $T_C/T_{C,max} = 1 - 82.6 \times (0.16 - \delta)^2$ [7].

Obviously, the BCS theory cannot be applied to explain superconductivity in HTSC since it gives an upper limit of $T_C$ at around 40K. So far, several microscopic models were suggested to account for high-Tc superconductivity, such as the RVB theory, Hubbard model, and t-J model. The RVB theory for HTSC was proposed by Anderson in 1987[8], but this model does not possess any anti-ferromagnetic long-range order. Soon after that, Anderson proposed the mean field theory based on the Hubbard model with a large U limit[8]. In the same year, Emery proposed a theory based on the three-band Hubbard model, where an anti-ferromagnetic long-range order can be obtained while electrons are half-filled[9], but this theory is too complicated. The physical picture in the Hubbard model is the existence of the empty, doubly-occupied and singly-occupied (spin up or down) states. In 1988, Zhang and Rice simplified the three-band Hubbard model to a sing-band t-J model in the large U limit[10], where the Hamiltonian consists of the kinetic energy term $t$ and the potential energy term $J$ describing the anti-ferromagnetic interaction. The physical picture in the t-J model is the existence of the empty and singly-occupied (spin up or down) states but the nonexistence of the doubly-occupied states. The t-J model can only explain some properties of HTSC, not all properties. In a word, there has not been a well-accepted theory that can explain superconductivity in HTSC since it was discovered 30 years ago. In this paper, a magnetic-interaction-induced theory for HTSC was presented, and we hope this theory can bring us a better understanding about superconductivity.

We will first introduce four idealized assumptions about this theory.

(1) Only a small number of electrons near the Fermi surface contribute to superconductivity in HTSC, and there are three types of electrons: the localized electrons $n_L$, normal conductive electrons $n_n$ and superconducting electrons $n_S$, and thus $n = n_L + n_n + n_S$ ($n$ is the electron density). The density of superconducting electrons $n_S$ is $\delta$ times the carrier density which is $n\delta$, i.e., $n_S = n\delta^2$, and $n_n = n\delta - n\delta^2$. And then $n_L = n - n\delta$ is the density of the localized electrons.

(2) Magnetic interactions between electron spins lead to Cooper pairs near the Fermi surface, and give rise to superconductivity; and all other kinds of interactions between the superconducting electrons can be neglected. This magnetic interaction energy is exactly the condensed energy which is given by the sum of the bonding energy of the Cooper pairs, i.e., $n_S\Delta$. An external magnetic field can destroy this magnetic interaction, break up Cooper pairs, and transform the superconducting states into normal states. Therefore, the maximum value of the condensed energy density equals the maximum energy density of the external magnetic field $B_C^2/2\mu_0$.

(3) There are different electronic states in HTSC, i.e., the on-site doubly-occupied, off-site doubly-occupied, singly-occupied (spin up or down) and empty states. Only the off-site doubly-occupied electrons can form the superconducting states (or Cooper pairs), while those on-site doubly-occupied electrons form the anti-ferromagnetic insulator states, and the singly-occupied (spin up or down) electrons form the normal states. All other possible physical states are neglected.

(4) The average kinetic energy of electrons (or holes) in HTSC complies with the equipartition theorem of energy. Based on the above theorem, the average kinetic energy of electrons is $ik_B T/2$ ($i = 1,2,3 \cdots$), where $i$ is the electronic degrees of freedom, and the maximum value of it is $ik_B T_C/2$. An increase in temperature can break up Cooper pairs, and thus destroy superconductivity. Therefore, the bonding energy of a Cooper pair is equal to $2\Delta = ik_B T_C$, and the superconducting energy gap is given by $\Delta = ik_B T_C/2$. In HTSC, Tc is approximately proportional to $\Delta$ with a relation of $\Delta = 4.8 k_B T_C/2$.

## 2. An effective Hamiltonian for HTSC

An approximate effective Hamiltonian to describe magnetic-interaction-induced super-conductivity in HTSC will be presented based on the assumptions above. All the electrons can be classified into three types: the localized electrons, normal conductive electrons and superconducting electrons.

First, the localized electrons with a density of $n_L = n - n\delta$ are the doubly-occupied electrons on the same lattice sites (on-sites) without any kinetic energy. The corresponding Hamiltonian can be written as

$$H_I = U \sum_i n_{i\uparrow} n_{i\downarrow} - J \sum_i n_{i\uparrow} n_{i\downarrow}. \tag{1}$$

There are a strong Coulomb interaction $U$ and an anti-ferromagnetic interaction $J$ for these localized electrons that make undoped copper oxides in the anti-ferromagnetic insulator states.

Second, the normal conductive electrons with a density of $n_n = n\delta - n\delta^2$ are those of the singly-occupied (spin up or down) electrons without any correlation on the different lattice sites. These electrons only have the kinetic energy without any potential energy, and the corresponding Hamiltonian is given by

$$H_{II} = t \sum_{i,j} (C_{i\uparrow}^+ C_{i\uparrow} + C_{j\downarrow}^+ C_{j\downarrow}). \tag{2}$$

where $t$ is the electron kinetic energy. The normal conductive electrons can be considered as free electrons, and they make doped-insulators metallic-like behaviour.

Third, the superconducting electrons with a density of $n_S = n\delta^2$ consist of the correlated doubly-occupied electrons on the different lattice sites (off-sites). The Cooper pair has the kinetic energy $2t$ and anti-ferromagnetic interaction potential $J$, and thus the corresponding Hamiltonian can be expressed by the off-site pairing operators

$$H_{III} = 2t \sum_{i,j} b_{ij}^+ b_{ij} - J \sum_{i,j} b_{ij}^+ b_{ij} \tag{3}$$

with $b_{ij}^+ = (C_{i\uparrow}^+ C_{j\downarrow}^+ - C_{i\downarrow}^+ C_{j\uparrow}^+)/\sqrt{2}$ [8]. These superconducting Cooper pairs can make the doped anti-ferromagnetic insulators into superconducting states. If the kinetic energy $2t$ of the Cooper pairs is less than the anti-ferromagnetic interaction potential $J$, these Cooper pairs move and form the superconducting states. On the other hand, if it is larger than the anti-ferromagnetic interaction potential $J$, the superconducting Cooper pairs will be broken up and converted into the normal conductive electrons.

Now we can get the total effective Hamiltonian of HTSC from Eqs. (1), (2) and (3)

$$H = U \sum_i n_{i\uparrow} n_{i\downarrow} - J \sum_i n_{i\uparrow} n_{i\downarrow} + t \sum_{i,j} (C_{i\uparrow}^+ C_{i\uparrow} + C_{j\downarrow}^+ C_{j\downarrow}) + 2t \sum_{i,j} b_{ij}^+ b_{ij} - J \sum_{i,j} b_{ij}^+ b_{ij}. \tag{4}$$

These Hamiltonians in Eq.(4), RVB, 2-D Hubbard and t-J models share commonalities as well as dissimilarities.

### 3. The superconducting transition temperature

Next we will derive a formula of Tc in HTSC based on the above model.

There are superconducting, normal conductive and localized electrons in HTSC, and only a small amount of electrons contributes to superconductivity. The number of superconducting electrons is maximum at $T = 0K$, but still significantly less than the total number of carriers in HTSC. The maximum density of superconducting electrons is $n_{S,max} = n\delta_{opt}^2 \neq n\delta_{opt}$. These superconducting electrons combine into Cooper pairs via magnetic interactions, and the formation of bound states will lower the energy of the system, leading to the superconducting states. With increasing temperature (or external magnetic field), this magnetic interaction in Cooper pairs will be destroyed, leading to a decrease of the number of superconducting electrons. While the temperature of the system increases to $T_C$ (or the external magnetic field reaches to $B_C$), all the Cooper pairs will be broken up and the number of superconducting electrons drops to zero. The energy density in a superconducting system is given by the sum of the electronic potential, the kinetic energy, and the energy of the external magnetic field

$$E = -n_{S,max}\Delta + \frac{B^2(T)}{2\mu_0} + n_{S,amx}\frac{i}{2}k_B T. \tag{5}$$

If $E \leq 0$, HTSC will be in the superconducting states, and thus we get $B \leq \sqrt{2\mu_0 n_{S,max}\Delta} = B_C$ at $T = 0K$ (or $T \leq (2\Delta/ik_B) = T_C$ at $B = 0T$). This is the critical magnetic field $B_C$ (or the critical temperature $T_C$). The condensed energy density in HTSC is maximum at $T = 0K$ and it equals the maximum energy density of the external magnetic field, i.e.,

$$n_{S,max}\frac{i}{2}k_B T_{C,max} = \frac{B_{C,max}^2}{2\mu_0}. \tag{6}$$

We can thus obtain the relation among $T_C$, the doping concentration $\delta$ and the critical magnetic field $B_C$ and the electron degrees of freedom $i$

$$T_{C,max} = \frac{1}{k_B\mu_0} \cdot \frac{B_{C,max}^2}{ni\delta_{opt}^2}. \tag{7}$$

The formula of $T_C$ in Eq. (7) does not contain the strong interaction $U$ and anti-ferromagnetic interaction energy $J$, then we will derive $T_C$ from the Hamiltonian in Eq.(4).

The total energy density of the superconducting electron system in the eigenstates is given by

$$E_T = (n - n\delta)\frac{U-J}{2} + (n\delta - n\delta^2)t + \frac{n}{2}\delta^2(2t - J)$$

$$= -n\frac{J}{2}\left(\delta - \frac{2t-U+J}{2J}\right)^2 + n\frac{J}{2}\left(\frac{2t-U+J}{2J}\right)^2 + n\frac{U-J}{2}. \tag{8}$$

Based on the above assumptions that the average kinetic energy of all the electrons follows the energy equipartition theorem, the total energy density is $E_T = nik_B T_C/2$. At $\delta = \delta_{opt} = (2t - U + J)/2J$, the maximum energy density can be obtained from Eq.(8),

$$E_{T,max} = n\frac{J}{2}\left(\frac{2t-U+J}{2J}\right)^2 + n\frac{U-J}{2} = n\frac{i}{2}k_B T_{C,max}. \tag{9}$$

Then the maximum value of $T_C$ is

$$T_{C,max} = \frac{J}{ik_B}\left(\frac{2t-U+J}{2J}\right)^2 + \frac{U-J}{ik_B}. \tag{10}$$

The relation between $T_C$ and the doping concentration $\delta$ can be derived from Eq.s (8) and (10),

$$\frac{T_C}{T_{C,max}} = 1 - \left[\left(\frac{2t-U+J}{2J}\right)^2 + \frac{U-J}{J}\right]^{-1}\left(\delta - \frac{2t-U+J}{2J}\right)^2. \tag{11}$$

At $U = 0.9865J$, $t = 0.15325J$, the optimal doping concentration is $\delta_{opt} = (2t - U - J)/2J = 0.16$, then Eq.(11) is changed into,

$$\frac{T_C}{T_{C,max}} = 1 - 82.64(\delta - 0.16)^2. \tag{12}$$

Eq.(12) is identical with the empirical formula of the correlation between $T_C$ and the doping concentration $\delta$, and thus can account for the phase diagram of HTSC.

Since Eq.(10) is a complicated expression for $T_C$ due to the strong interaction $U$ and anti-ferromagnetic interaction $J$, the localized electrons can be considered as the background, i.e., the first term in Eq.(8) is neglected ($U - J = 0$), thus, we obtain

$$E_T = (n\delta - n\delta^2)t + \frac{n}{2}\delta^2(2t - J) = n\delta t - \frac{n}{2}\delta^2 J. \tag{13}$$

At $\delta = \delta_{opt} = t/J$, $T_C$ in Eq.(10) can be simplified by

$$T_{C,max} = \frac{1}{ik_B}\left(\frac{t}{J}\right)^2 J = \frac{\delta_{opt}^2}{ik_B}J. \tag{14}$$

And the relation between $T_C$ and the doping concentration $\delta$ in Eq.(11) can also be simplified into

$$\frac{T_C}{T_{C,max}} = 1 - \left(\frac{J}{t}\right)^2\left(\delta - \frac{t}{J}\right)^2. \tag{15}$$

At the optimal doping $\delta_{opt} = t/J = 0.16$, Eq.(15) can be changed into

$$\frac{T_C}{T_{C,max}} = 1 - 39.06(\delta - 0.16)^2. \tag{16}$$

The $T_C$ values calculated from Eq.(14) for YBa$_2$Cu$_3$O$_{6.15}$, La$_2$CuO$_4$, SmOFeAs are 92.8K ($i = 2$, $\delta = 0.16$, $J\delta = 100$meV), 40.3K ($i = 1$, $\delta = 0.16$, $J = 136$meV), and 58K ($i = 2$, $\delta = 0.2$, $J\delta = 50$meV), respectively. These results are in good agreement with the experimental values of 92K, 40K and 54K. Moreover, it was estimated that $T_C$ in the slab superconductor is higher than that in the bulk for the same material, and $T_C$ for SmOFeAs can be up to 116K.

## 4. Conclusions

Based on four idealized assumptions we proposed a microscopic theory of magnetic-interaction-induced pairing in superconductivity of HTSC. An effective Hamiltonian to describe HTSC was suggested; and a parabolic relation between $T_C$ and the doping concentration $\delta$ was derived and the results can be used to explain the phase diagram for HTSC. It was found that, $T_C$ is correlated to the anti-ferromagnetic interaction energy $J$ (or critical magnetic field $B_C$) and the freedom degrees of electrons $i$. It was thus calculated that $T_C$ is 92.8K for YBa$_2$Cu$_3$O$_{6.15}$, 40.3K for La$_2$CuO$_4$, and 58K for SmOFeAs, which are in good agreement with the experimental values. It was estimated that, $T_C$ in the slab HTSC is higher than that in the bulk, and $T_C$ for SmOFeAs can be up to 116K.